\documentclass[la4paper,12pt,oneside]{article}
\usepackage{hyperref}
\usepackage{url}
\usepackage{graphicx}
\usepackage{amsmath}
\usepackage{amssymb}
\usepackage[noline]{algorithm2e}
\usepackage{geometry}
\usepackage{longtable}
\usepackage{amsthm}

\usepackage{pgf,tikz}
\usepackage{tkz-euclide}
\usepackage{tkz-graph}

\usetikzlibrary{calc}
\usetkzobj{all}

\usepackage{pgfplots}
\newlength\figureheight 
\newlength\figurewidth 

\newtheorem*{lem*}{Lemma}
\newtheorem*{pro*}{Proof}
\newtheorem*{thm*}{Theorem}
\newtheorem*{defn*}{Definition}
\newtheorem*{rem*}{Remark}
\newtheorem*{app*}{Appendix}
\newtheorem*{problem*}{Problem}
\newtheorem*{assump*}{Assumption}
\newtheorem*{example*}{Example}
\newtheorem*{cor*}{Corollary}
\newtheorem*{nota*}{Notation}

\usepackage{overpic}

\date{}

\title{A Note on Properties of Discrete Composition Operators \footnote{This research was supported by the ERC Starting Grant No. 307047 (COMET).}}

\author{Klaus Glashoff\footnote{Universita della Svizzera italiana}, Claus Peter Ortlieb\footnote{University of Hamburg, Germany}}

\begin{document}
\maketitle

Let $\mathcal{M}=\{p_1,p_2,...,p_m\}$ and $\mathcal{N}=\{q_1,q_2,...,q_n\}$ denote finite sets with $m\leq n$, and let $\tau : \mathcal{M} \rightarrow \mathcal{N}$ be an \emph{injection}, i.e. $\tau(p_i)=\tau(p_j) \Leftrightarrow i=j.$ We define an injective map $\pi :\{1,2,...,m\}\rightarrow \{1,2,...,n\} $ by
$$ \pi (i)= j  \Longleftrightarrow \tau(p_i)=q_j.$$
\begin{rem*}
In case $m=n$, $\pi : \{1,2,...,m\} \rightarrow \{1,2,...,m\}$ is a \emph{permutation}.
\end{rem*}
Let now $P: \mathbb{R}^\mathcal{N} \rightarrow \mathbb{R}^\mathcal{M}$ denote the \emph{discrete composition operator}, defined by
$$  (Pf )_i= f_{\pi (i)}$$
which maps real functions on $\mathcal{N}$ to real functions on $\mathcal{M}$ (\emph{pullback}). Because of $|\mathcal{M}|=m$, and $|\mathcal{N}|=n$, P can be represented by a real $m \times n$ matrix which, with  slight abuse of notation, is also denoted by $P$.

\begin{defn*}
An $m\times n-$ matrix $P$ with $m\leq n$ generated by an injective map $\pi :\{1,2,...,m\}\rightarrow \{1,2,...,n\} $ is called a \emph{discrete composition matrix}.
\end{defn*}

The aim of the following theory is to present properties of discrete composition matrices  which are useful for the treatment of such matrices in the field of numerical computation of shape correspondences (see \cite{Nogneng2017}).

\begin{defn*}
Let $P$ be a real $m\times n-$ matrix ($m,n\geq 1$). $P$ is called \emph{row-permutation-like}, if
\begin{enumerate}
  \item{All matrix elements are  0 or 1}
  \item{Each row contains at most one 1}
\end{enumerate} 
\end{defn*}

It is rather obvious that the matrix $P$ representing a discrete composition operator, is row-permutation-like. It contains a $1$ at the i-th row and the j-th column, if $\tau (p_i) = q_j$ (which is equivalent to $\pi (i) = j$), and all the other elements of $P$ are $0$. In case of $m=n$, the composition matrix is a permutation matrix, which follows from the injectivity of $\pi$.

We are now going to present two different characterizations of row-permutation-like matrices.  In the following, $f.g$ denotes the element-wise multiplication of vectors, and $<v,w>$ the scalar product of a row vector $v$ with a column vector $w$.  $diag(f)$ denotes the diagonal matrix containing the components of $f$ on the diagonal.

The following result holds for \emph{arbitrary} $m, n \geq 1$. Theorems of this type for more general composition operators on spaces of measurable functions, can be found in \cite{singh1993composition}.
\begin{thm*}
Let $P$ denote a real matrix of dimension $m\times n$. The following statements are equivalent.
\begin{enumerate}
\item{$P$ is row-permutation-like}
\item{$P(f.g)=(Pf).(Pg)$ for all $f,g \in \mathbb{R}^n$}
\item{$Pdiag(f)=diag(Pf)P$ for all $f\in \mathbb{R}^n $}
\end{enumerate}
\end{thm*}

\begin{pro*}
$2.\rightarrow 1.$: Let $e_j, e_k$ denote the j-th and k-th unit vector in $\mathbb{R}^n$, respectively. We consider two cases, first the case $j=k$. Let $p_{i,.}$ be an arbitrary row of P; it follows from 2. that  $p_{i,j}=p_{i,j}^2$, which shows the first property of a permutation-like matrix, namely, that $p_{i,j}$ is $0$ or $1$. This is true for each $i, 1\leq i \leq m$ and $j, 1\leq j \leq n$. Then assume that $j\neq k$, which by 2. implies, for each $i, 1\leq i \leq m$ (i.e., for each row of $P$), $0=p_{i,j}*p_{i,k}$. This implies that no row of $P$ has a $1$ at different places (column indices). Thus $P$ is row-permutation-like.

$1.\rightarrow 2.$: Let $p$ denote an arbitrary row vector of $P$. Then, by 1., $p^T$ is either the null vector or a unit vector of $\mathbb{R}^n$ which implies $<p,(f.g)> =<p, f ><p,g>$. As this is true for all rows of $P$, the assertion of 2. is proven.

$2. \leftrightarrow 3.$: $P(f.g)=(Pf).(Pg)$ for all $f,g \in \mathbb{R}^n$ is equivalent to  $Pdiag(f)g=diag(Pf)Pg$ for all $f,g \in \mathbb{R}^n$, which in turn is equivalent to 3. \flushright $\square$
\end{pro*}

Matrices $P$ generated from discrete composition maps $\pi :\{1,2,...,m\}\rightarrow \{1,2,...,n\} $ are, as has been said above, row-permutation-like, but not every row-permutation-like matrix can be generated by such a $\pi$. For example, the null-matrix is row-permutation-like, and this applies also to a matrices like
$$\left(\begin{array}{cccc}
1 & 0 & 0 & 0\\
1 & 0 & 0 & 0\\
0 & 0 & 0 & 0
\end{array}\right)$$
which is row-permutation-like but not a discrete composition matrix.

In addition to being row-permutation-like, discrete composition matrices are also \emph{column-permutation-like}. This means that a composition matrix $P$ has at most one $1$ in each column. This follows from the fact that the map $P$ is generated by a \emph{map}  $\pi$ which cannot send one index $i$ to two different images. (In the example matrix above, the index $1$ is sent to $1$ \emph{and} $2$.) 
Thus we define

\begin{defn*}
Let $P$ be a real $m\times n-$ matrix ($m,n\in \mathbb{N}$). $P$ is called \emph{column-permutation-like}, if
\begin{enumerate}
  \item{All matrix elements are  0 or 1}
  \item{Each column contains at most one 1}
\end{enumerate}
\end{defn*}

Of course a matrix $P$ is column-permutation-like if and only if its transpose $P^T$ is row-permutation-like. This leads to the following characterization of column-permutation-like matrices.

\begin{thm*}
Let $P$ denote a real matrix of dimension $m\times n$. The following statements are equivalent.
\begin{enumerate}
\item{$P$ is column-permutation-like}
\item{$P^T(f.g)=(P^Tf).(P^Tg)$ for all $f,g \in \mathbb{R}^m$}
\item{$diag(f)P=Pdiag(P^Tf)$ for all $f\in \mathbb{R}^m $}
\end{enumerate}
\end{thm*}

(3. follows by transposition of $P^Tdiag(f)=diag(P^Tf)P^T$, which is just property 3. of the characterization of row-permutation-like matrices, applied to $P^T$).
As we know that every composition matrix $P$ is as well row-permutation-like as also column-permutation-like, we get the following
\begin{thm*}
Let $P$ be an $m\times n-$ discrete composition matrix $(m\leq n)$. Then
\begin{enumerate}
 \item{$P(f.g)=(Pf).(Pg)$ for all $f,g \in \mathbb{R}^n$  and $P^T(f.g)=(P^Tf).(P^Tg$ for all $f\in \mathbb{R}^m $}
 \item{$Pdiag(f)=diag(Pf)P$ for all $f\in \mathbb{R}^n $ and $diag(f)P=Pdiag(P^Tf)$ for all $f\in \mathbb{R}^m $}
\end{enumerate}
\end{thm*}

The converse of this theorem is not true as can be seen by the following example of a row- and column-permutation like matrix which is \emph{not} a discrete composition matrix (because of the last row of zeros):
$$\left(\begin{array}{cccc}
1 & 0 & 0 & 0\\
0 & 1 & 0 & 0\\
0 & 0 & 0 & 0
\end{array}\right)$$

This shows that, for a complete characterization of discrete composition matrices, an additional condition is necessary which guarantees that $rank(P) = m$; for example the condition that all row sums are $1$. This then leads to the following 

\begin{cor*}
An $m\times n-$ matrix with $m\le n$ is a discrete composition matrix if and only if the following three conditions are fulfilled:
\begin{enumerate}
 \item{$P(f.g)=(Pf).(Pg)$ for all $f,g \in \mathbb{R}^n$  and $P^T(f.g)=(P^Tf).(P^Tg)$ for all $f\in \mathbb{R}^m $}
 \item{$Pdiag(f)=diag(Pf)P$ for all $f\in \mathbb{R}^n $ and $diag(f)P=Pdiag(P^Tf)$ for all $f\in \mathbb{R}^m $}
 \item{All row-sums are 1}.
\end{enumerate}
\end{cor*} 

\emph{Acknowledgement}. The first author thanks \emph{Dorian Nogneng}, Ecole Polytechnique Palaiseau, France, for illuminating discussions on this subject induced by the  paper  \cite{Nogneng2017}.

\bibliographystyle{plain}
\bibliography{biblio}

\end{document}